# On-demand acoustic shaping of Mossbauer gamma-ray photons

I. R. Khairulin[1,2], and Y. V. Radeonychev[1,2]

[1] *Institute of Applied Physics of the Russian Academy of Sciences,*
*46 Ulyanov Street, Nizhny Novgorod, 603950, Russia,*
[2] *N.I. Lobachevsky State University of Nizhny Novgorod,*
*23 Gagarin Avenue, Nizhny Novgorod 603950, Russia*

We propose a technique that makes it possible to transform the intensity of a quasi-monochromatic single-photon wave packet, emitted by a radioactive Mössbauer γ-ray source, into a sequence of short bursts with an arbitrary number of bursts, including a single burst. In addition, the technique allows one to individually and independently control, on demand, the moments of the burst appearance, as well as the peak intensity, duration and shape of each burst in the sequence. The technique is based on the transmission of Mössbauer (recoilless) photons through a resonantly absorbing medium, which is rapidly displaced at some moments of time relative to the source (or vice versa) along the photon propagation direction at a distance less than the photon wavelength, and returned to its original position. The burst durations can be comparable to the duration of the single-photon pulses produced by synchrotrons but have the controlled spectral-temporal characteristics. We show that the proposed technique can be implemented on the basis of currently available equipment with use of 14.4-keV recoilless photons, emitted by $^{57}$Co source, and $^{57}$Fe absorber, which opens up prospects for its applications in Mössbauer spectroscopy and x-ray quantum optics.

## I. INTRODUCTION

Methods for generating coherent electromagnetic radiation in the form of short pulses with the controlled spectral and temporal characteristics in various spectral ranges from terahertz to x-rays have been intensively developed due to numerous applications in science and technologies. The topical problems of quantum optics, including quantum information processing and quantum communications, stimulate the search for ways to control properties of single photons. Presently, there are lots of methods for controlling optical photons, which commonly use driving lasers and different materials including specially designed micro- or nanoscale spatial structures. However, it is much more difficult to control x-ray photons with energy of tens of keV. This is mainly due to the short wavelength of such photons. Indeed, optical photons are controlled through their interactions with atomic electrons whose optical properties are changed by laser fields. On the contrary, the main effects of interaction of angstrom-wavelength photons with electrons atoms are ionization and recoil of the latter. In addition, sufficiently high-intensity and spectrally bright coherent sources in the x-ray range are currently absent. As a result, most approaches well suited for controlling optical and infrared photons are currently not feasible for hard x-ray photons.

At the same time, high-energy photons can get into resonance with quantum transitions of atomic nuclei and efficiently interact with them without recoil (Mössbauer interaction) similar to the interaction of optical photons with atomic electrons. In addition, unlike atomic transitions, recoilless nuclear resonances are usually almost naturally broadened, and their spectral width can be very small. For example, the 14.4-keV Mössbauer (recoilless) nuclear transitions in $^{57}$Fe nuclide normally have a few orders of magnitude narrower linewidths at room temperature (several megahertz) than transitions of bound electrons in atoms (several gigahertz), corresponding to enormous Q-factor on the order of $10^{12}$. Together with the high density of the nuclei in a solid (normally from a few units by $10^{21}$ up to $10^{23}$ cm$^{-3}$) this allows for a large resonant optical depth with a small physical length of the absorber (an *e*-fold attenuation of 14.4-keV radiation is achieved in a $^{57}$Fe foil of about 70 nm thickness).

Very high frequencies of the keV nuclear transitions makes the Doppler effect very efficient tool for controlling the frequencies of these transitions. For example, motion of $^{57}$Fe absorber at a constant velocity of 0.17 mm/s along the photon propagation direction shifts the position of its 1.13 MHz-wide spectral line, centered at wavelength of 0.86 Å, by 2 MHz. These properties of nuclei underlie Mössbauer spectroscopy. They also open up prospects for developing compact photonic devices for x-ray quantum optics as well as give rise to novel mechanisms and methods for acoustic control of high-energy photons via their resonant interaction with atomic nuclei ([1-33] and references therein).

The main sources of single photons in the tens of keV range are radioactive Mössbauer isotopes and synchrotrons after spectral narrowing of extremely broadband x-ray pulses emitted from the synchrotron storage ring. Both types of sources normally emit, on average, less than one photon in a certain direction during the characteristic photon emission time. Therefore, spectral and temporal characteristics of the emitted electromagnetic field are obtained as a result of multiply repeated detection of isolated photons. If the beginning of the detection of each photon is locked to the moment from which this photon can appear, then the entire set of measurements forms the corresponding characteristics of a single-photon coherent wave packet or, equivalently, of a single-photon pulse. The time and frequency dependences of the photon detection probability, measured using this technique, correspond to the temporal and spectral dependences of intensity of the coherent single-photon pulse. The time dependence of the intensity of the single-photon wave packet is the photon waveform.

In the case of a synchrotron source, the beginning of photon detection is locked to the deterministic moments of emission of the original synchrotron radiation (SR) pulse. Single-photon x-ray pulses produced by synchrotrons have a short duration of about 100 ps and a huge spectral width of several terahertz due to widely spread carrier frequencies of photons even after strong monochromatization of SR beam. If x-ray photons are monochromatized at an energy of 14.4 keV, their spectrum and waveform can be modified by means of transmitting photons through iron or stainless-steel single or two foils enriched with resonant $^{57}$Fe nuclei [23-28]. In this case, the transmitted 100-ps single-photon pulse is followed by a long tail of the coherent field (with a half-height duration of about 100 ns) forward-scattered by $^{57}$Fe nuclei (nuclear coherent response to the incident pulse) [23-28]. The spectrum and waveform of this tail were modified either by the rapid displacement of one of the foils by half the wavelength of the field at certain moments of time [26-28], or by harmonic ultrasonic vibration of the foil with an amplitude of half the wavelength of the photon [24,25] along the photon propagation direction. Both types of foil motion caused phase modulation of the coherent forward-scattered single-photon field due to the Doppler effect, which led to a strong change in its spectrum [27,28], but to a relatively small change in the tail of the photon waveform [24-26]. Along with that, the picosecond pulse ahead, much stronger than the subsequent tail, can neither be removed nor controlled.

The radioactive sources emit recoilless γ-ray photons stochastically as a result of spontaneous radioactive decay. A number of isotopes such as $^{57}$Co, $^{67}$Ga, $^{129m}$Te, demonstrate possibility to a cascade radioactive decay with emission of photon sequences. For example, the $^{57}$Co nucleus emits a photon with the energy of 122 keV followed by the 14.4-keV photon (Fig. 1). In this case, the 14.4-keV single-photon wave packet is obtained by locking the beginning of the 14.4-keV photon detection to the moment of detection of the 122-keV photon. The detection of the 122-keV photon heralds that since that moment the 14.4-keV photon can appear. In the case of a single-line radioactive source, a photon waveform has a stepwise front edge and a long exponentially decreasing tail due to the Lorentz-like shape of the near-naturally broadened radiative nuclear transition. For example, the half-height duration of coherent single-photon pulse from $^{57}$Co source is 97 ns.

A long duration of recoilless photons from the radioactive source makes it possible to significantly change both the spectrum and waveform of the emitted photon by means of transmitting photons through a resonantly absorbing medium under conditions when the source is rapidly (much faster than the duration of the single-photon pulse) displaced relative to the absorber periodically or at certain times ([1-4,10-17,16-21] and references therein). Similar to the synchrotron source, the displacement causes the phase shift of the coherent forward-scattered field, produced by the absorber nuclei, relative to the phase of the incident field. However, unlike the synchrotron source, the forward-scattered field interferes with the field of the source. This interference can be controlled via shifting the phase between the incident and forward-scattered fields.

Using this method, the intensity of the 14.4-keV photon stream emitted by $^{57}$Co source was strongly modulated when passed through a motionless optically thick resonant $^{57}$Fe absorber. Short and intense bursts were obtained via a stepwise displacement of the source by the photon wavelength $\lambda \approx 0.86$Å, as well as via oscillatory displacement of the source with an amplitude of about $4\lambda$ and a period close to the single-photon pulse duration [3]. Measurement of the 14.4-keV photon waveform (performed with locking the 14.4-keV photon detection to the detection of the 122-keV photon) showed that an abrupt displacement of the source by $\lambda/2$, $\lambda$, and $3\lambda/2$ at a certain time within the photon duration can lead to the transformation of its exponential waveform with a stepwise front edge into two short bursts [4]. One burst is the unremovable Sommerfeld-Brillouin precursor and the second is a short burst produced due to the source displacement. Subsequently, it was shown that in the case of abrupt displacement of the absorber by $\lambda/2$, the intensity of the produced burst can be increased by optimizing the absorber optical depth and the frequency detuning between the source and absorber spectral lines [6], as well as by use of several sequentially arranged absorbers [7,8].

Another technique of Doppler control of the spectrum and waveform of the recoilless high-energy photons was proposed in [11-13], implemented in [11,14-16], and theoretically developed in [19-22,29-31]. This technique is based on vibration of the absorber or source with period much shorter than duration of the single-photon pulse and with the same amplitude for all nuclei within the photon beam. The latter was achieved by a piston-like vibration of a stainless-steel foil. The vibration amplitude was usually less than half the wavelength of photon. The phase-modulated forward-scattered field of the high-frequency vibrating absorber has a spectrum in the form of a comb of well-resolved phase-matched components. In this case, the quasi-monochromatic field of the incident photon interferes with only one component of the spectral comb. A proper choice of the interfering component and the adjustment of its frequency to resonance with the source spectral line results in the transformation of the exponential waveform of the incident photon into a regular sequence of short pulses. At the optimal amplitude of the absorber vibration, the duration and repetition period of the pulses are determined by the absorber vibration frequency: the higher the vibration frequency, the shorter both the pulses and the repetition period.

The maximum peak intensity of the pulses produced by the high-frequency harmonic vibration of the absorber is less than the maximum peak intensity of the pulses produced by the stepwise displacement of the absorber under optimal conditions. As shown in [31], there is an optimal anharmonic regular reciprocating displacement of the absorber maximizing the peak intensity of the produced pulses. In the implementation, this displacement coincides with the stepwise displacement of the absorber by the photon wavelength $\lambda$ for producing each pulse.

In this paper, we propose a novel technique of absorber displacement that allows one to independently control the temporal characteristics of an individual recoilless γ-ray photon emitted by a radioactive source [34]. It is based on a *rapid reciprocating* displacement of the resonant absorber along the propagation direction of photons *by a variable distance less than the wavelength of the photon field*. With the example of $^{57}$Co source, and $^{57}$Fe absorber we show that this type of motion makes it possible to (i) produce a single-photon wave packet in the form of variable number of short bursts including a single burst (except for the Sommerfeld-Brillouin precursor); (ii) individually adjust the peak intensity of each burst in the sequence; (iii) individually adjust the duration of each burst in the sequence; (iv) adjust the individual time interval between the adjacent bursts in the sequence; (v) produce on-demand an individual shape (e.g., increasing exponent, triangle, or meander) of each burst in the sequence.

The paper is organized as follows. In Section II, using an example of $^{57}$Co single-photon source, we show how the exponentially decreasing photon waveform transforms into a short burst when the photon passes through an optically deep recoilless resonant $^{57}$Fe absorber rapidly reciprocating displaced by half the photon wavelength at certain moment along the photon propagation direction. We discuss the physical mechanism of this transformation and the basic characteristics of the produced burst. In Sec. III, we show how the characteristics of the produced single or several bursts in the intensity of the single-photon wave packet are independently controlled by the characteristics of the absorber displacements. In Section IV we summarize the results.

## II. ABSORBER DISPLACEMENT BY HALF OF γ-RAY PHOTON WAVELENGTH

Let us consider the 14.4-keV photons, produced by a radioactive Mössbauer source $^{57}$Co, which propagate through a resonant $^{57}$Fe absorber (Fig. 1). As mentioned above, in the direction of detection, the radioactive source emits normally single photons separated in time. In the following, we assume that the 14.4-keV photons are repeatedly detected with the start of detection locked to the moment of registration of the 122-keV photon. In this case, we obtain the characteristics of the single-photon wave packet (single-photon pulse).

The intensity of 14.4-keV single-photon pulse emitted by the $^{57}$Co source, $I_s(t,t_0)$, is proportional to the time dependence of the probability for detecting the 14.4-keV photon per unit of time (the normalized coincidence count rate of the 14.4-keV photons), starting from the moment of detection of the 122-keV photon, $t_0$, which is usually set to zero. Before the absorber, the intensity can be written as

$$I_s(t,t_0) \propto \theta(t-t_0)\exp\left[-(t-t_0)\Gamma_s\right], \qquad (1)$$

where $\theta(\tau)$ is the unit step function and $\Gamma_s$ is the decay rate of the state $|b\rangle$ (Fig. 1).

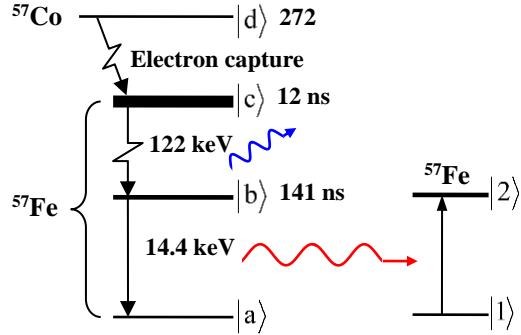

**Fig. 1.** The energy diagrams of the $^{57}$Co source (left side) and the $^{57}$Fe absorber (right side). The $^{57}$Co nucleus radioactively decays via electron capture transforming to the $^{57}$Fe nucleus in the excited state $|c\rangle$, which then decays with lifetime $1/\Gamma_c \simeq 12\,\text{ns}$ to the state $|b\rangle$ via the emission of a 122-keV photon. Subsequently, the $^{57}$Fe nucleus decays with lifetime $1/\Gamma_s \simeq 141\,\text{ns}$ from the state $|b\rangle$ to the state $|a\rangle$ emitting a 14.4-keV photon. The moment of emission of the 122-keV photon determines the moment $t_0$ of the formation of the state $|b\rangle$, to which the beginning of detection of the 14.4-keV photon can be locked for measuring characteristics of the 14.4-keV single-photon wave packet.

According to [1-8,10-16,19-21,35-37], at the entrance to the absorber, the corresponding electric field of the 14.4-keV single-photon wave packet can be represented in the form:

$$E_s(t,t_0) = E_0 \theta(t-t_0) e^{-(i\omega_s+\gamma_s)(t-t_0)+i\varphi_0}, \qquad (2)$$

where $E_0$ is the field amplitude, $\gamma_s = \Gamma_s/2$ is the half-width of the photon spectral contour, $\omega_s$ is the carrier frequency of the field corresponding to the wavelength $\lambda_s \approx 0.86$Å, $\varphi_0$ is the random initial phase of the field.

The 14.4-keV single-photon wave packet propagates through a resonant Mössbauer absorber $^{57}$Fe, which moves as a whole (at the scale of the beam cross section) along the photon propagation direction according to a displacement function $S_{shift}(t)$. In practice, the resonant absorber is a several micron thick stainless-steel foil, enriched with resonant $^{57}$Fe nuclei. The foil is attached to a piezoelectric transducer constituting a polyvinylidene fluoride (PVDF) film. The piezoelectric PVDF film can change its thickness depending on the magnitude and sign of the applied electric voltage. This change in thickness can be made the same within the cross section of the photon beam, which provides piston-like (as a whole) displacement of the irradiated part of the stainless-steel foil [14-16,18,24,25].

Similar to references [14-16,20,22,29-31], the description of photon propagation through a moving absorber is simplified and become more clear in a frame of reference, co-moving with the absorber. In this reference frame, the resonant nuclei are motionless, while the incident field becomes phase-modulated due to the Doppler effect:

$$E'_{in}(t,t_0) = E_0 \theta(t-t_0) e^{-(i\omega_s+\gamma_s)(t-t_0)+ik_s S_{shift}(t)+i\varphi_0}, \qquad (3)$$

where prime indicates the moving reference frame and $k_s = 2\pi/\lambda_s$. Then the photon electric field at the exit from the resonant absorber with the Mössbauer (optical) thickness $T_a$ and the half-width of the absorption line $\gamma_a$ can be calculated as the convolution integral of the incident field (3) and the absorber response function $a(t)$ in the form [1-7,37,38]:

$$E'_{out}(t) = \int_{-\infty}^{\infty} a(t-\tau) E'_{in}(\tau) d\tau, \qquad (4)$$

with

$$a(t) = e^{-T_e/2} \left[ \delta(t) - T_a \gamma_a e^{-(i\omega_a+\gamma_a)t} \theta(t) J_1\left(\sqrt{2T_a\gamma_a t}\right) / \sqrt{2T_a\gamma_a t} \right]. \qquad (5)$$

In (5), $\delta(t)$ is the Dirac delta function, $J_n(p)$ is the Bessel function of first kind of the $n$-th order ($n \in \mathbb{Z}$), $\omega_a$ is the frequency of the resonant transition $|1\rangle \leftrightarrow |2\rangle$ of the $^{57}$Fe absorber (Fig. 1), which can differ from the central frequency of the source, $\omega_s$, due to isomeric shift or Doppler shift produced by a motion of the absorber relative to the source with a constant velocity, and $T_e$ is the exponent factor of photoelectric absorption. In what follows, we assume for simplicity that the photon field (3) is in resonance with the absorber, $\omega_s = \omega_a \equiv \omega$, and the widths of the spectral contours of the source and absorber are the same, $\gamma_s = \gamma_a \equiv \gamma$. Then Eq. (4) can be rewritten as:

$$E'_{out}(t,t_0) = E_0 \theta(t-t_0) e^{-(i\omega+\gamma)(t-t_0)+i\varphi_0} e^{-T_e/2} \left[ e^{ik_s S_{shift}(t)} + A_{AR}(t,t_0) \right], \quad (6)$$

where

$$A_{AR}(t,t_0) = -T_a \gamma \int_{t_0}^{t} \frac{J_1\left(\sqrt{2T_a\gamma(t-\tau)}\right)}{\sqrt{2T_a\gamma(t-\tau)}} e^{ik_s S_{shift}(\tau)} d\tau. \quad (7)$$

According to (6) (except for the term $e^{-T_e/2}$), the photon field at the exit from the absorber is the sum of the incident field (3) (the first term in (6)), and the absorber response or, equivalently, the coherent field forward-scattered by the absorber nuclei (the second term in (7)).

The intensity of the single-photon pulse at the exit from the absorber directly follows from (6). It can be written in the form

$$I_{out}(t,t_0) = I_0 \theta(t-t_0) e^{-T_e - 2\gamma(t-t_0)} \left\{ \left(1 - |A_{AR}(t,t_0)|\right)^2 + 4|A_{AR}(t,t_0)| \cos^2\left(\frac{k_s S_{shift}(t) - \arg(A_{AR}(t,t_0))}{2}\right) \right\} \quad (8)$$

where $I_0 = cE_0^2/(8\pi)$ and $A_{AR} = |A_{AR}|\exp[i\arg(A_{AR})]$. As follows from (8), the intensity of the single-photon wave packet does not depend on the reference frame.

If the absorber is at rest, $S_{shift}(t) \equiv 0$, then the amplitude of the forward-scattered field (7) is antiphase with respect to the incident field and can be written in the form

$$A_{AR}^{(rest)}(t,t_0) = \left[1 - J_0\left(\sqrt{2T_a\gamma(t-t_0)}\right)\right] e^{i\pi}. \quad (9)$$

Destructive interference of the forward-scattered field with the incident field, described by Eq.(6), leads to attenuation of the single-photon field in the absorber. The characteristic rate of this attenuation can be estimated as $1/\tau_a$ where $\tau_a$ is the characteristic response time of the resonant optically thick absorber assumed as the interval, during which the Bessel function $J_0(p)$ changes from unit at $t=t_0$ to zero at $\sqrt{2T_a\gamma\tau_a} \approx 2.4$ in equation (9). Thus, $\tau_a \approx 2.9/(\gamma T_a)$. At $t-t_0 = \tau_a$, the amplitude of the forward scattered field becomes equal to the amplitude of the incident field, which leads to the cancellation of the output field. In the optically thick absorber, the attenuation rate $1/\tau_a$ of the single-photon field (6),(8) can greatly exceed the attenuation rate $\gamma_s$ of the incident photon field (2), which is called the speed-up effect [37]. The oscillatory time dependence of the transmitted photon waveform, described by the function $J_0^2\left(\sqrt{2T_a\gamma(t-t_0)}\right)$ in (8), is called dynamical beats [35,37].

Let us consider the model dependence of the absorber displacement on time in the form of short piecewise linear function. Namely, the displacement starts at the moment $t_{start} > t_0$, then during the interval $\Delta t$ the absorber is displaced from the source by the amplitude $\Delta z$, and then returned to its original position during the same interval,

$$S_{shift}(t) = \begin{cases} 0, t < t_{start}, \\ \dfrac{\Delta z}{\Delta t}(t - t_{start}), t_{start} \leq t < t_{start} + \Delta t, \\ \Delta z - \dfrac{\Delta z}{\Delta t}(t - t_{start} - \Delta t), t_{start} + \Delta t \leq t < t_{start} + 2\Delta t, \\ 0, t \geq t_{start} + 2\Delta t. \end{cases} \quad (10)$$

As follows from (7), if the total duration of the absorber displacement, (10), $\Delta t_{total} = 2\Delta t$, is much shorter than the characteristic response time of the resonant optically thick absorber, $\Delta t_{total} \ll \tau_a$, the forward-scattered field amplitude (7) is approximately equal to

$$A_{AR}(t,t_0) \approx A_{AR}^{(rest)}(t,t_0) - T_a\gamma\Delta t_{total} \times \begin{cases} 0, t_0 \leq t < t_{start}, \\ \dfrac{1}{2}F\left[(t-t_{start})/\Delta t_{total}\right], t_{start} \leq t < t_{start} + \Delta t_{total}, \\ F(1)J_1\left(\sqrt{2T_a\gamma(t-t_{start})}\right)/\sqrt{2T_a\gamma(t-t_{start})}, t \geq t_{start} + \Delta t_{total}, \end{cases} \quad (11)$$

where

$$F(x) = \int_0^x \exp\left[ik_s S_{shift}\left(t_{start} + \xi\Delta t_{total}\right)\right]d\xi. \quad (12)$$

Obviously, if $\Delta t_{total} \ll 2.9/(\gamma T_a)$ and $t_{start} > t_0 + 2.9/(\gamma T_a)$, one can consider that $A_{AR}(t,t_0) = A_{AR}^{(rest)}(t_{start},t_0)$, and $\left|A_{AR}^{(rest)}(t_{start},t_0)\right|$ is on the order of unit during the entire time of the absorber displacement, $t_{start} \leq t \leq t_{start} + \Delta t_{total}$. Thus, if $\Delta z = \lambda_s/2$ in (10), then during the absorber movement, the phase of the incident field in the reference frame of the absorber (the first term in brackets in (6)) changes from 0 to $\pi$ when moving forward, and then from $\pi$ to 0 when returning to the initial position. At the moment $t^* = t_{start} + \Delta t$, that corresponds to the absorber shift by the half-wavelength of photon, $\lambda_s/2$, the incident field becomes in-phase with the coherently scattered field, and their constructive interference leads to a strong burst of the intensity of a single-photon wave packet.

If, as in studies [4-7], the absorber is stopped at the moment $t_{start} + \Delta t$ ($\Delta t \ll \tau_a$ corresponding to the stepwise-like displacement) and $\Delta z = \lambda_s/2$, then, according to (6), the resulting field in the absorber reference frame, is

$$E'_{out}(t,t_0) \approx E_0 \theta(t-t_0) e^{-(i\omega+\gamma)(t-t_0)+i\varphi_0} e^{-T_e/2} \begin{cases} 1 + A_{AR}^{(rest)}(t,t_0), \ t_0 \leq t < t_{start}, \\ e^{i\pi(t-t_{start})/\Delta t} + A_{AR}^{(rest)}(t,t_0), \ t_{start} \leq t < t_{start} + \Delta t, \\ e^{i\pi} + A_{AR}^{(rest)}(t,t_0) + 2e^{i\pi} A_{AR}^{(rest)}(t,t_{start}), \ t \geq t_{start} + \Delta t, \end{cases} \quad (13)$$

where the third term in the bottom line describes the nuclear response on the incident phase-reversed field together with the forward-scattered field generated at $t < t_{start} + \Delta t$. The intensity of the field (13) is plotted in Fig. 2a, and the short burst was called Gamma echo [4,5]. As follows from (13) and (9), the duration of the produced short burst is on the order of $1/(\gamma T_a)$. It can be much shorter than the duration of the incident single-photon pulse if the absorber is optically thick. However, the duration of the produced burst is determined by the properties of the absorber and cannot be controlled.

On the other hand, if, after shifting by half the photon wavelength during the interval $\Delta t$, the absorber is rapidly returned to its original position during the same interval according to the displacement function (10), then the phase of the incident field relative to the coherently scattered field changes from $\pi$ to 0. At the moment when the absorber stops in its original position, the incident field again becomes antiphase to the almost unchanged coherently scattered field (9) and, hence, will be attenuated, as before the absorber displacement. Thus, if the absorber is displaced in accordance with the displacement function (10), a short intense burst with a duration $\Delta t_{total} = 2\Delta t$ (from the beginning to the end) will be formed in the photon waveform at the exit from the absorber (Fig. 2b) due to the short-term constructive interference of the incident and coherent forward-scattered field. In this case, the duration of the burst can be much shorter than the duration of Gamma echo due to the stepwise absorber displacement. In addition, the duration of the burst is completely controlled by the displacement duration.

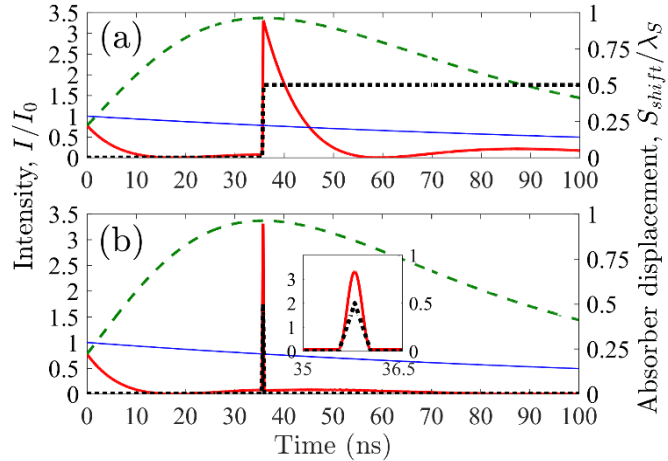

**Fig. 2.** Time dependences of the normalized intensity of the 14.4-keV single-photon pulse at the exit from the resonant nuclear absorber according to (8),(7) (red solid lines and left axes). The absorber is displaced according to the displacement function (10) along the direction of photon propagation (black dotted lines and right axes). Panel (a) corresponds to the case of Gamma echo [4,5] when the absorber is displaced at the moment $t_{start} = 35.56$ ns by half the wavelength of the 14.4-keV photon ($\Delta z = \lambda_s/2 = 0.43$ Å) during time $\Delta t = 0.25$ ns, followed by a stop. Panel (b) corresponds to the case when the absorber is displaced at the same moment $t_{start} = 35.56$ ns by half the wavelength of the resonant photon, followed by a return to its original position during the time $2\Delta t = 0.5$ ns. In both panels, the green dashed line describes the dependence (14) of the peak intensity of the produced burst on the moment of its formation, the thin blue line is the waveform of the incident photon. The intensity is normalized to the peak intensity of the incident single-photon pulse. The parameter values are $t_0 = 0$, $T_a = 45.7$, $T_e = 5.6 \times 10^{-3} T_a$ corresponding to the stainless-steel foil $Fe_{70}Cr_{19}Ni_{11}$ in which 90% of the Fe nuclei are the resonant nuclide $^{57}Fe$.

According to (8), the maximum peak intensity of the produced burst is realized when $k_s S_{shift} + \arg A_{AR} = \pi$ independently of the form of the absorber reciprocating displacement, $S_{shift}(t)$.

However, it essentially depends both on the Mössbauer thickness of the absorber and on the moment at which the burst achieves its maximum, $t_{max}$. Using the approximation $A_{AR}(t,t_0) = A_{AR}^{(rest)}(t,t_0)$ and substituting it into (8), one can obtain an approximate dependence of the peak intensity of the burst on the Mössbauer thickness, $T_a$, and the moment of formation of the burst maximum, $t_{max}$,

$$I_{out}^{max}(T_a,t_{max}) = I_0 \theta(t_{max}) e^{-T_e - 2\gamma t_{max}} \left[ 2 - J_0\left(\sqrt{2T_a \gamma t_{max}}\right) \right]^2, \tag{14}$$

where $t_{max} = t_{start} - t_0 + \Delta t$. For a commercially available Mössbauer absorber in the form of a stainless-steel foil $Fe_{70}Cr_{19}Ni_{11}$ in which 90% of the Fe nuclei are the resonant nuclide $^{57}Fe$, the estimate $T_e = 5.6 \times 10^{-3} T_a$ is applicable. Then, similarly to [20,21], one can find the maximum possible intensity of the burst, $\left(I_{out}^{max}\right)_{opt} \approx 3.37 I_0$, as well as the moment $t_{max}^{(opt)} \approx 35.8\,\text{ns}$, at which it is achieved, and the optimal optical thickness of the absorber, $T_a^{(opt)} \approx 360 \gamma t_{max}^{(opt)} \approx 45.7$ (see the green dashed lines in Fig. 2) corresponding to the physical thickness of the stainless-steel foil of about 4.5 μm. In order to produce this most intense burst, the duration of the absorber displacement, according to the condition $\Delta t_{total} \ll 2.9/(\gamma T_a)$, should be $2\Delta t \ll 17.8\,\text{ns}$. The duration of the burst will also be much shorter than 17.8 ns.

The sawtooth time dependence of the absorber displacement (10) with amplitude $\Delta z = \lambda_s/2$ can be implemented by one cycle of harmonic vibration in the form

$$S_{shift}(t) = \frac{\lambda_s}{2} \sin^2\left(\pi \cdot \frac{t - t_{start}}{2\Delta t}\right)\left[\theta(t - t_{start}) - \theta(t - t_{start} - 2\Delta t)\right]. \tag{15}$$

In this case, the corresponding frequency of oscillation with maximum displacement of 0.43 Å (half the wavelength of 14.4-keV photon) should be much higher than 56 MHz. Such a vibration of a stainless-steel foil can be implemented using a PVDF transducer [39]. It should be noted that the higher the frequency and the magnitude of the voltage applied to the PVDF film, the more difficult is to provide the piston-like displacement of its surface with the attached stainless-steel foil within the beam cross section. At a given frequency of the applied voltage, the displacement of the PVDF film surface becomes more piston-like if its thickness changes with a smaller amplitude. Taking this into account, the stainless-steel foil displacement (15) can be realized with a change in the PVDF-film thickness by 0.215 Å (quarter the wavelength of 14.4-keV photon) from its undisturbed magnitude if the PVDF film is pre-compressed to a constant value of 0.215 Å before its thickness is rapidly changed. This technique corresponds to the displacement function (15) shifted by $\Delta z = \lambda_s/4$ (Fig. 3), i.e.,

$$S_{shift}(t) = -\frac{\lambda_s}{4} + \frac{\lambda_s}{2} \sin^2\left(\pi \cdot \frac{t - t_{start}}{\Delta t_{total}}\right)\left[\theta(t - t_{start}) - \theta(t - t_{start} - 2\Delta t)\right]. \tag{16}$$

This technique also makes it easier to suppress the relaxation oscillations of the PVDF film when it is forced to stop.

It should also be noted that for a given value $\gamma$ of the nuclear transition decay rate, the condition $\Delta t_{total} \ll 2.9/(\gamma T_a)$ for obtaining the strongest burst in a given absorber can be met with a slower displacement of the absorber if it has a smaller optical thickness, $T_a$. In this case, a weaker and slower developing nuclear response will result in a longer burst with peak intensity less then $3.37 I_0$. As an example, let us consider the stainless-steel foil, used in [14], with the natural abundance (about 2%) of $^{57}Fe$, having the physical thickness of 25 μm that corresponds to the optical thickness $T_a \approx 5$ and the photoelectric absorption factor $T_e \approx 1.27$ [40]. If it is harmonically displaced according to (16) over the time $\Delta t_{total} = 33$ ns (corresponding to the vibration frequency of 30 MHz), the burst of about 16 ns duration at half-maximum (which is about $\Delta t_{total}/2$) with amplitude $I_{out}^{max} \approx 0.35 I_0$ can be obtained (Fig. 3a). In this case, $2.9/(\gamma T_a) \approx 162$ ns, therefore the condition $\Delta t_{total} \ll 2.9/(\gamma T_a)$ is not perfectly met. However, one can see that both the duration and shape of the burst approximately repeat the displacement function (16). The large distortion in the burst shape is only near its maximum. This is due to a slow change in the position of the absorber near the return point. The corresponding change in the phase of the incident field in the absorber reference frame is also slow compared to $\tau_a$, so that the nuclear response has time to become partially antiphase during the slow motion of the absorber.

In the case of relatively slow displacement of the absorber, $\Delta t_{total} > 2.9/(\gamma T_a)$, a short burst with the half-maximum duration of about $\Delta t_{total}/2$ will also be produced (Fig. 3b). In this case, the nuclear response has time to follow a relatively slow phase change of the incident field and partially maintains the destructive interference with it most of the time. As a result, the produced burst with a duration slightly shorter than $\Delta t_{total}/2$ has a lower peak intensity and more distorted shape.

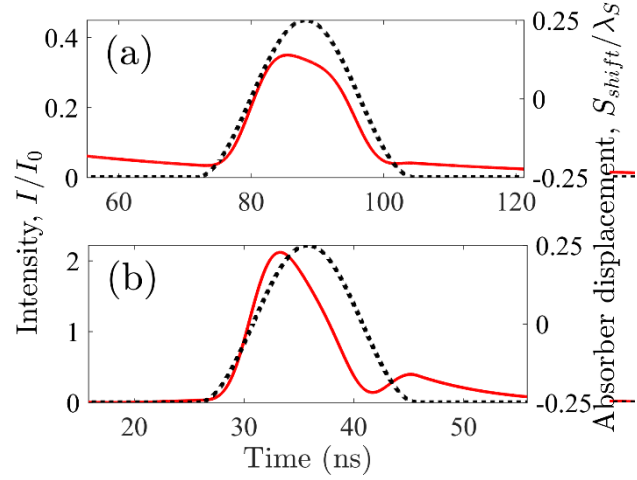

Fig. 3. Time dependences of the normalized intensity of the 14.4-keV single-photon pulse (red solid lines and left axes) at the exit from the resonant nuclear absorber, according to (8),(7). The absorber is displaced according to the displacement function (16) along the direction of photon propagation (black dotted lines and right axes). In both panels, the intensity is normalized to the peak intensity of the incident single-photon pulse and $t_0 = 0$. In panel (a), $t_{start} = 71.72$ ns, $\Delta t_{total} = 33$ ns, $T_a = 5$ ($2.9/(\gamma T_a) \approx 162$ ns $\approx 4.9 \Delta t_{total}$), $T_e = 1.27$. In panel (b), $t_{start} = 25.81$ ns, $\Delta t_{total} = 20$ ns, $T_a = 45.7$ ($2.9/(\gamma T_a) \approx 17.8$ ns $= 0.89 \Delta t_{total}$), $T_e = 5.6 \times 10^{-3} T_a$.

## III. CONTROLLED SHAPING THE γ-RAY PHOTON WAVEFORM

As shown above, in the case of a rapid reciprocating motion of the absorber, the duration of the displacement straightforwardly determines the duration of the burst produced in the photon waveform. The burst appears at the moment when the absorber begins to move, and the intensity of the burst drops to zero at the moment when the absorber stops in its original position. Obviously, this cycle can be repeated an arbitrary number of times with arbitrary intervals between the cycles and with arbitrary duration of each cycle. As a result, one can transform the exponentially decreasing waveform of the incident photon into an arbitrary sequence of short bursts, in which the number of bursts, intervals between the adjacent bursts, and the duration of each burst can be independently controlled on demand (Fig. 4).

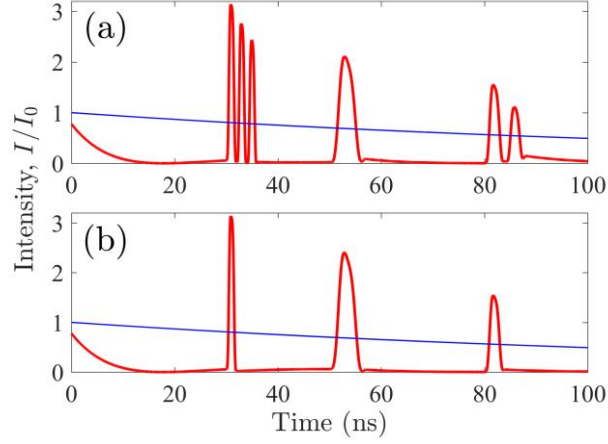

**Fig. 4.** Time dependences of the normalized intensity (8),(7) of the 14.4-keV single-photon pulse at the exit from the $^{57}$Fe resonant absorber with $T_a = 45.7$, $T_e = 5.6\times10^{-3}T_a$, which is displaced along the photon propagation direction according to the displacement function (16) thrice, once and twice during 6 ns, 7 ns, and 8 ns at the moments of 30 ns, 50 ns and 80 ns, respectively, on panel (a), and once during 2 ns, 7 ns, and 4 ns at the moments 30 ns, 50 ns, and 80 ns, respectively, on panel (b). The thin blue line is the waveform of the incident photon. The intensity is normalized to the peak intensity of the incident single-photon pulse.

Now let us consider one of the short reciprocating displacements of the absorber in the case when the displacement has an arbitrary form, while the entire duration of the displacement, $\Delta t_{total}$, is much shorter than the characteristic response time of the absorber, $\Delta t_{total} \ll 2.9/(\gamma T_a)$. As follows from (8) and (11), in this case, the shape of the photon intensity at the exit from the absorber is determined only by the form of the absorber displacement function via the term $k_s S_{shift}(t)$. Indeed, according to (11), for $t_{start} \leq t \leq t_{start} + \Delta t_{total}$, $t_{start} > t_0 + 2.9/(\gamma T_a)$, the approximation $A_{AR}(t,t_0) \approx A_{AR}^{(rest)}(t_{start},t_0) = -A$ ($A > 0$) is valid, similar to the sawtooth displacement function (10). Hence, in this interval, the intensity (8) takes the form

$$I_{out}(t,t_0) = I_0\theta(t-t_0)e^{-T_e-2\gamma(t-t_0)}\left\{(1-A)^2 + 4A\sin^2\left[\pi S_{shift}(t)/\lambda_s\right]\right\}, \quad t_{start} \leq t \leq t_{start} + \Delta t_{total}, \qquad (17)$$

where $(1-A)^2 \ll 1$. As can be seen in (17), the shape of each short burst in the photon waveform at the exit from the absorber, $I_{out}(t,t_0)$, maps the shape of the absorber displacement, $S_{shift}(t)$. The mapping function is $\sin^2\left[\pi S_{shift}(t)/\lambda_s\right]$. Some examples are shown in Fig. 5.

As also follows from (17), in the case of an arbitrary shape of the absorber displacement, if its maximum, $|S_{shift}(t)|_{max} \equiv S$, meets the condition $S \leq \lambda_s/2$, then the maximum of the absorber displacement determines and controls the peak intensity of the produced burst. Indeed, according to (17),

$$I_{out}^{peak}(S) = I_0 e^{-T_e-2\gamma(t_{start}-t_0+\Delta t)}\left\{(1-A)^2 + 4A\sin^2\left(\pi S/\lambda_s\right)\right\} \approx I_{out}^{peak}(\lambda_s/2)\sin^2\left(\pi S/\lambda_s\right), \qquad (18)$$

This is illustrated in Fig. 5a,b for the sawtooth displacement function (10). For example, if $S = \lambda_s/4$ (Fig. 5b), then $I_{out}^{peak}(\lambda_s/4) \approx 0.5 I_{out}^{peak}(\lambda_s/2)$ If the amplitude of the absorber displacement is $\lambda_s/2 < S \leq \lambda_s$, then the two-hump burst will be produced (Fig. 5c) due to an oscillatory dependence of $\sin^2\left(\pi S_{shift}(t)/\lambda_s\right)$ in this case.

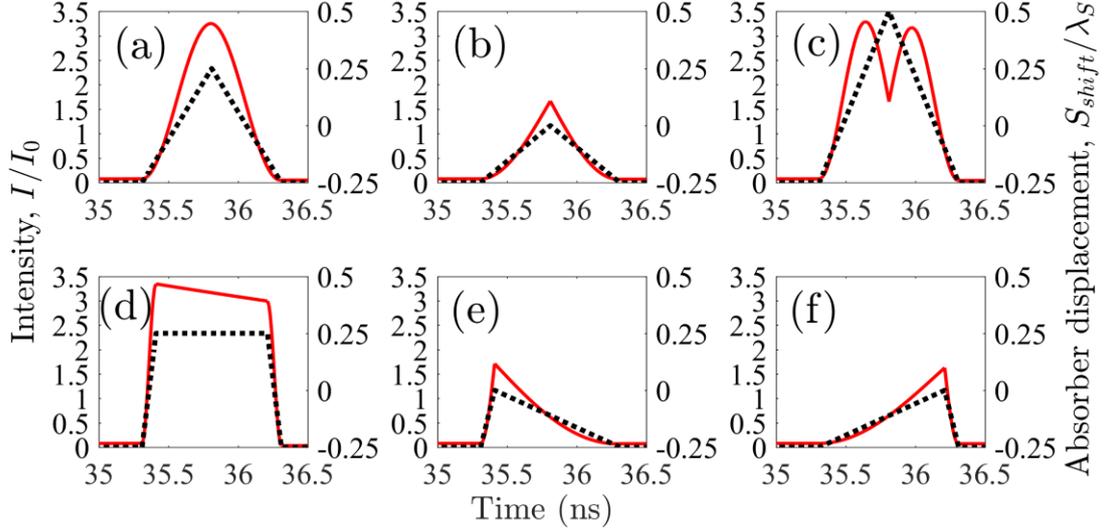

**Fig. 5.** Time dependences of the normalized intensity (8),(7) (red solid lines and left axes) of the 14.4-keV single-photon pulse at the exit from the resonant $^{57}$Fe absorber with $T_a = 45.7$, $T_e = 5.6 \times 10^{-3} T_a$, which is displaced along the photon propagation direction according to different model displacement functions (black solid lines and right axes). The intensity is normalized to the peak intensity of the incident single-photon pulse.

In Fig. 5d, the absorber doesn't move during 0.8 ns. In this interval, the nuclear response according to (13) is described by the function $2e^{i\pi} A_{AR}^{(rest)}(t, t_{start})$ (see (9)), which determines the slope of the plateau-like top of the burst. The mapping function $\sin^2\left[\pi S_{shift}(t)/\lambda_s\right]$ is applicable only for domains of rapid displacement of the absorber.

One can show that in the case of a slower displacement of the absorber with a smaller optical thickness, so that the condition $\Delta t_{total} \ll 2.9/(\gamma T_a)$ is met, the photon waveform can be controlled by the absorber displacement function in the same way as above. For example, in the case of the absorber with $T_a = 5$ used in [14], its reciprocating displacement during time $\Delta t_{total} \approx 30$ ns with the displacement functions plotted in Fig. 5 leads to similar bursts of about 30 ns duration (measured from the beginning to the end of the burst) and lower than in Fig. 5 peak intensities.

## IV. CONCLUSION

In this paper, we have proposed a technique that makes it possible to transform the intensity of a quasi-monochromatic single-photon wave packet emitted by a radioactive Mössbauer γ-ray source into a sequence of short bursts with an arbitrary number of bursts, including a single burst. The technique also allows one to individually and independently control, on demand, the moments of the burst appearance, as well as the peak intensity, duration and shape of each burst in the sequence. The technique is based on the transmission of Mössbauer (recoilless) photons through a resonantly absorbing medium, which is rapidly displaced at some moments of time relative to the source (or vice versa) along the photon propagation direction by a distance less than the wavelength of the photon field, and returned to its original position. When the absorber (or source) begins to move, the destructive interference between the incident field and the field coherently scattered by the nuclei, is converted into constructive interference due to the Doppler effect. This leads to a sharp increase in intensity. When the absorber (or source) rapidly returns to its original position, destructive interference is restored, which forces the intensity quenching. As a result, a short burst in the transmitted field intensity is appeared. The burst begins with the beginning of the absorber (or source) displacement and finishes at the moment when the absorber (or source) stops in its original position. The peak intensity and shape of the burst map the amplitude and time dependence of the absorber (or source) displacement. The burst durations can be several orders of magnitude shorter than the duration of the incident single-photon pulse and can be comparable to the duration of single-photon pulses produced by synchrotrons. At the same time, unlike the synchrotron pulses, these bursts have the controlled temporal characteristics. We have shown that the proposed technique can be implemented with use of 14.4-keV recoilless photons, emitted by $^{57}$Co source, and $^{57}$Fe absorber. This opens up prospects for full acoustic control of the temporal characteristics of recoilless photons for applications in Mössbauer spectroscopy and x-ray quantum optics.

**ACKNOWLEDGMENTS**

This work was supported by the Center of Excellence "Center of Photonics" funded by the Ministry of Science and Higher Education of the Russian Federation, contract 075-15-2020-906.